# Optimal path selection in Graded network using Artificial Bee Colony algorithm with Agent enabled Information

Kavitha Sooda
Advanced Networking Research Group, RIIC, DSI
and Asst. Professor, Dept. of CSE
Nitte Meenakshi Institute of Technology
Bangalore - 560064, India
e-mail: kavithasooda@gmail.com

T. R. Gopalakrishnan Nair
ARAMCO Endowed Chair- Technology, PMU, KSA
Advanced Networking Research Group
VP, Research, Research and Industry Incubation Center
(RIIC), Dayananda Sagar Institutions,
Bangalore - 560078, India
e-mail: trgnair@ieee.org, www.trgnair.org

*Abstract*— **In this paper we propose a network aware approach for routing in graded network using Artificial Bee Colony (ABC) algorithm. ABC has been used as a good search process for optimality exploitation and exploration. The paper shows how ABC approach has been utilized for determining the optimal path based on bandwidth availability of the link and how it outperformed non graded network while deriving the optimal path. The selection of the nodes is based on the direction of the destination node also. This would help in narrowing down the number of nodes participating in routing. Here an agent system governs the collection of QoS parameters of the nodes. Also a quadrant is synthesized with centre as the source node. Based on the information of which quadrant the destination belongs, a search is performed. Among the many searches observed by the onlooker bees the best path is selected based on which onlooker bee comes back to source with information of the optimal path. The simulation result shows that the path convergence in graded network with ABC was 30% faster than non-graded ABC.**

## I. INTRODUCTION

In the recent past, Swarm intelligence has been applied for studying collective behavior of known population. It has become a research interest to many research scientists of related fields in recent years. The Internet in present scenario has confronted with problems like security, scalability, performance due to rapid increase in the end-users and new service demands. Internet test-beds are required to support reliable end-to-end research for federation and management of Internet. Here routing and forwarding has become the core problem for delivery of data from one node to another. Current routing strategy whether it is wired or wireless routing protocols, are based on some predetermined calculation or prior knowledge of the connection between the nodes. This information was simply based on the request packets sent. Generally for any intelligent action [1] to take place a collective behavior of the topology needs to be studied prior to the path determination process. Also there exists a need for self organizing the topology when changes occur. This has led to the introduction of Swarm intelligence, a nature inspired process to handle this kind of situation [2-3]. Here the agents locally collect information based on some predefined parameter [4]. Later they all communicate with each other globally to arrive at the desired result. Because of its self adaptive property, swarm intelligence has gained importance in many field of research.

The remainder of the paper is organized as follows: Section 2 outlines the overview of Network Awareness and Graded network. Section 3 deals with ABC algorithm. Section 4 presents the details of the mathematical model. Section 5 consists of performance results. Finally, the last section concludes the paper.

## II. NETWORK AWARENESS AND GRADED NETWORK

### A. Network Awareness

Network awareness [5] is much dependent on querying the distributed system state space such as the scenario of Internet. It deals with supplementing the knowledge of network characteristics while deciding on the network locations where the query operators are executed. There exists a need for querying widely in distributed system that enables intelligent data acquisition from the nodes which can participate in network routing. Existing querying approaches are suitable for small-scale systems and it fails when the network grows. The network awareness approach helps in one way the implementation of querying process in a large distributed environment. In general, the network operating principle is largely based on the characteristics like topology and bandwidth. In this paper we incorporate the network aware management strategy. It utilizes the network characteristics to identify better nodes where the operator can approach or query the node to identify suitable situation. The network characteristic is dealt based on the QoS metric such as the bandwidth. The algorithm, initially assume that any given node is a feasible candidate for obtaining the optimal path the forward push of messages. Decision is taken on nodes, based on the query operator which will bifurcate the nodes in to feasible the non-feasible nodes. Next the neighbor of the previous feasible node is considered and the process is repeated until the destination node is reached. Here the network aware algorithm is based on exploration of the neighbor nodes as illustrated in Fig. 1.

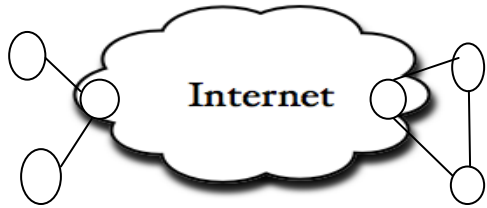

Figure 1. Network Awareness Routing

Thus the environment awareness [5] can extract the nodes which are feasible to propagate the data in an efficient manner. This leads to an improved performance due to reduction in the search space and provides feasible path for the data to reach the desired destination in an efficient manner.

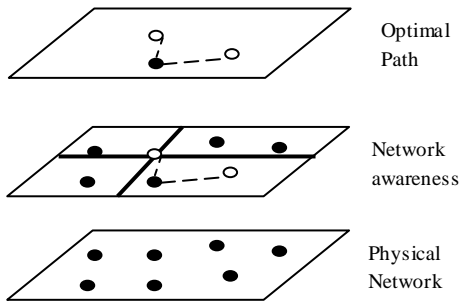

Figure 2. Network Awareness algorithm

Here in this paper, the network awareness algorithm i.e, level-1 approach of grading algorithm is applied on a physical network where the nodes which participate in routing are available as in [4]. Now a quadrant is designed where the quad is drawn with source as the centre point as shown in Fig. 2. Now the nodes which belong to the destination quadrant will participate in routing. Thus the optimal path is determined assuming sufficient nodes are availble for access and selection in the selected quadrant of interest.

### B. Graded network

The authors in [4] have discussed much on the graded network. The Grade value estimation method for implementing intelligent routing in the autonomic network is the core focus of research and is the thrust. Grade is like an index; it is made available everywhere and routing will much depend on it. It helps building the quality of the router, which is in fact triggered by the knowledge of the environment. The router must be an intelligent entity because it performs different operations using the environment information. It depends on input, output, load and resource availability. The graded network is setup with six QoS parameters. These are considered at two levels; level-1 and level-2. *Level-1* is applied region-wise and the goal is to achieve favorable routing based on selected attributes. The values obtained from *Level-1* must be able to eliminate the non-production node. Non-production nodes are those that come in the pitfall of congestion and possess less resource availability. These nodes must be identified by the algorithm that defines the gradient from most non-productive to productive nodes in a homogeneous network. This is identified by assigning a grade value from -3 to +3, i.e. it signifies the productivity value of the node. At this point, we are able to calibrate the routing process region-wise. The algorithm requires proactive decision making on the output obtained. This is because many paths exist to reach the destination, and we have to choose the most optimal path. Once the gradient value has been calculated, it can be made available as pervasive information packets to all the other nodes to obtain the optimal path constitution.

Now the graded function, i.e., *Level-2*, considers the calculated values as its input of *level-1*. The output of this function defines the route availability for the set of nodes considered. This shall be calculated for all the available paths leading towards the destination node. The mean value of the gradient in the grade function shows the success level of operation of the network which is obtained by the ABC algorithm.

### III. BEE COLONY ALGORITHMS

In recent years swarm intelligence has been the focus of interest in research. Swarm intelligence is about developing an algorithm or a distributed problem solving device inspired by the collective behavior of social insect colonies and other animals. Swarm in general refers to any restrained collection of interacting agents or individuals. The classical examples which depict the behavior of swarm are bees swarming around their hive. Similar architecture which mimic the swarming approach are the ant colony, immune system, bird flocking and the fish schooling [6].

### A. Artificial Bee Colony(ABC) for network routing

Artificial Bee colony algorithm is developed by Karaboga in 2005 [7], which were motivated by the intelligent behavior of the bees. It is an optimization tool, where search is performed in a distributed environment. Here the artificial bee fly find the food source based on experience or their nest mates and adjusts their position. Some artificial bees (scouts) choose the food source randomly without experience. If the nectar amount of the new food source is better, then the bees memorize this position and forget the previous food source position. Thus ABC performs a local search with onlooker and employed bees, with global search methods managed by the onlooker and scout bees, attempting to balance exploration and exploitation process.

In ABC, the search space is exploited by the employed bees and onlooker bees and the scout bee does the exploration. Here the recruitment rate represents the measure of how quickly the swarm locates and exploits new food source. The survival of the bees here depends on how quickly they are able to discover optimal or feasible solutions. In Artificial Bee colony algorithm the initial food source is equal to the number of employed bees.

In this paper we apply these principles to solve routing problems optimally. They are used in the model to test and develop forward path progress. The nectar collected as food from the neighboring nodes is used as the parameter in equivalence to the bandwidth available at the forward point at the evaluation time point of the virtual network under consideration. When the messages are in different state of dispatch requirement, general algorithm restore to geographical connectivity mainly table driven algorithm. In a highly exploded model of network upcoming future limitations of getting stuck with geographical information need to be avoided and the message need to be distributed based on the intelligence of the router and the information collected from the environment such as available bandwidth at very forward node angular positioning of forward vector, QoS parameters like jitter, delay, etc. In the model the employed bees are randomly assigned to the food source for exploitation. In every iteration, every employed bee determines a food source in the neighborhood and evaluates its nectar amount using fitness function. The $i^{th}$ food source position is represented by $N_i \in (n_1, n_2, ......n_e)$. Here $\epsilon$ represents the feasibly connectable nodes. $F(n_i)$ determines the total nectar amount possibly accessible from $N_i$. After watching the dance of the employed bees, the onlooker bee goes to the region of food source at $N_i$ with probability $p_i$ defined as,

$$p_i = \frac{F(N_i)}{\sum_{k=1}^{f_s} F(n_k)} \qquad (1)$$

where $f_s$ is the total number of food sources. The numerator represents the nectar amount at node $N_i$ and denominator represents the total nectar amount accumulated from current node to all its neighbours. Here the dance represents the amount of free available bandwidth. The more is the bandwidth availability, the more the bees are flying towards the node.

To derive the intermediate paths, let $F_{max\,1}$ be the first neighbour of the current node. This function derives the filtered maximum value among all the neighbours from the current node we are looking from. Here = symbol represents the updates on intermediate paths obtained from the current node. The onlooker finds the neighborhood food source of $N_i$ and is represented using following equation where the + symbol represent the establishment of the intermediate path from current node to the node where maximum bandwidth to the destination direction is available,

$$N_i(n_i+1) = C_i(n_i) + F_{max\,1}^{w}(C_i(n_i)) \qquad (2)$$

The next intermediate node selection takes place by the set of onlooker bees. The onlooker bees' number depends on the number of neighbours each node posses which is represented by w. The selection of the neighbour from the current intermediate node is represented by $F_{max\,1}^{w}(C_i(n_i))$. The current intermediate node is represented by $C_i(n_i)$. $F_{max\,1}^{w}$ returns the node address which would have the highest bandwidth among all the neighbors of the current intermediate node. The result of this is concatenated with current node to determine the next intermediate path. Now $C_i(n_i)$ is updated as,

$$C_i(n_i) = N_i(n_i+1) \qquad (3)$$

This process is repeated until destination is reached or no link exists further to explore.

In this paper, when the system is setup a random activity shall start occupying certain bandwidth of the link. Here packets sent are of fixed size, i.e, 250 KB. Each node has 1MB of buffer space where the incoming packets are stored. The time taken to send one packet is one second from each node. The bandwidth utility on each link is monitored. A decision shall be taken whether to consider the link or not based on a threshold bandwidth. If the link has bandwidth above a threshold value then the link can participate in routing of the packets. This selection is applied after the quadrant structure is obtained.

Here the routers selected are the ones which are above the threshold bandwidth and the messages can transfer easily. The network awareness here is about identifying the nodes which satisfy QoS metric. This makes the routing decision to take place easily.

After the neighbours have been determined the fitness value is calculated. If this new fitness value is better than the old value then the bees move to this new food source otherwise it remains in the same node. When all employed bees have finished the search for food source it shares this information with the onlooker bees. Now onlooker bees select these food sources with a probability, $p_i$ as given by (1) which represents the best bandwidth on the link between $N_i(t)$ and $N_i(t+1)$.

Pseudo code on application of ABC which is applied for network routing is as follows:

Initialize the nodes and bandwidth randomly.
The employed bees are initially the number of neighbors to source
REPEAT
- The employed bees are moved onto their paths and fitness is calculated.
- The employed bees report to the onlooker bees about the information collection.
- Now selection is made where onlooker bees should move which is based on maximum bandwidth.
- If good node (means no link with sufficient bandwidth) not obtained scout bees (different quadrant search gets initiated) are used for searching.

- Memorize the best node found so far.

UNTIL (termination criteria satisfied)

Fig. 3 illustrate the flow chart of the implementation of the ABC algorithm for network routing. Here the populations of solutions are randomly initialized and evaluated to test to see whether it is above the threshold bandwidth value. The employed bees initially will be initialized to the number of neighbors of the source [8-9].

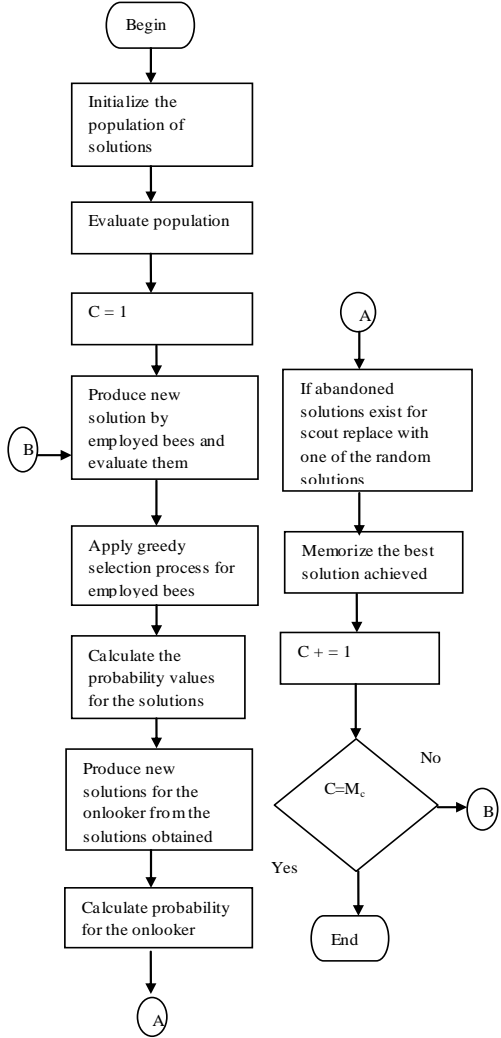

Figure: 3 ABC Algorithm for network routing

Here 'C' represents the cycle or the count for the performance of the ABC algorithm and $M_c$ represents the maximum cycle i.e., the total number of times ABC repeats.

The bee search for food source is to maximize the ratio E/T. Here E represents the amount of nectar food discovered and T represents the time spent for foraging. As far network is concern T is not permitted above a certain level for communication efficiency. And also foraging will take place not at very instant but at a selected interval of time which is again determined by the stability of the network in that particular region.

## B. Fitness function

The simulation topology is generated randomly. Here relaxation is imposed on the number of onlooker bees. The number of onlooker bees is equal to the number of neighbors to the source node who belong to the quadrant of the destination node. Here the fitness value is calculated on nodes which only belong to the quadrant. The nodes which satisfy the threshold bandwidth values are the ones which get selected and participate in routing which indicates the fitness value. Further the authors have planned to implement multi-objective parameters solvable by the fitness function [10].

## IV. MATHEMATICAL MODEL

Here we have proposed random network model with channel level stochastic Markov process for realizing the link capacity. Here every cell is considered to be a Jackson network [11]. Here links are interconnected by set of n links, L={1,2,3, …,n}, where every link has the same capacity, l ϵ L. let the flow arrival rate between any two nodes, say *s* and *e* be represented by $\lambda_{se}$. Here we assume that each flow transmits packet at fixed unit rate with mean $\mu^{-1}$, along the route it is assigned.

The traffic intensity of the link can be calculated as,

$$\text{Traffic Intensity} = \frac{Ps \cdot Tl(t)}{Ab} \quad (4)$$

Here Ps indicates the packet size, Ab indicates the available bandwidth and the traffic load on the link l ϵ L at time t is denoted by $T_l(t)$. It represents the total number of flows currently routed across it. The flow arrival rate to link l is denoted by $\gamma_l$. This flow represents a part of the flow arrival between *s* and *e* that is routed to link l. Here the flow departure rate is proportional to the current link load $T_l(t)$ and is given by $T_l(t) * \mu$. The packet size for the simulation is fixed as well as the available bandwidth on the link.

The updates on the link is done every 30*t sec, which would represent the delay involved in building a grade network. Here the model is developed based on the given information on traffic load and flow rate. Here we assume that the traffic arrivals during [0, t] are routed based on the link states $T_l(0)$ available at time 0. The link state $T_l(t)$ tracks to the following differential equations which derives the total traffic arrival rate $\gamma_l$ to link l.

$$\frac{d\, Tl(t)}{dt} = (\gamma_l - T_l(t) * \mu) \quad (5)$$

This is of the form,

$$\frac{dy}{dx} + Py = Q$$

Therefore the Integrating Factor(IF) is,

Y(IF) = ∫Q(IF)dt +c

Hence eq 5 becomes,

$$T_l(t) = T_l(0)\, e^{-\mu t} + \gamma_l / \mu [1 - e^{-\mu t}] \quad (6)$$

where $T_l(0)$ is the link state information at time interval [0,t]. This value is subtracted from link capacity to obtain the remaining fraction of the bandwidth available.

$B_a = L_c - T_l(0);$ (7)

Using Eq (6) and Eq (7) we can determine the traffic intensity from Eq. (4). Equation (7) helps is the calculation of whether congestion has occurred or not.

The model is designed based on the hexagon structure with nodes and link generated at random. In each of the cell structure, the jobs arrive from outside which are Poisson in process with rate $\alpha > 0$. Each arrival is independently routed to node j with probability $p_j \geq 0$ and $\sum_{j=1}^{k} P_{ij} = 1$. Here k represents the total number of nodes that a packet which has arrived from node i and can go through. Upon service completion at node i, a job may go to another node j with probability $p_{ij}$ or leave the cell with probability $q_i = 1 - \sum_{j=1}^{k} P_{ij}$. This is similar to the Jackson open network and mapped to Eq. (6), for calculating the load on link l.

Thus the determination of the link capacity helps to identify where congestion has occurred or not. The agent here is used for aiding the grade network with relevant information when required. For instance at the first level of grading the parameters which are looked for are Bandwidth availability, congestion level check, delay, node density and resource allocation. All these five parameters are taken care by the agent. For instance the bandwidth availability is determined by Eq. (7) which in turn helps to check whether congestion has occurred or not. Delay is obtained by Eq. (4). Network node density is number of packets arriving to the node and resource allocation is randomly assigned.

We then apply local algorithm at level-2 trying to achieve global optimization, which is referred to as constrained optimization. Here we applying the probabilistic reasoning approach to derive optimal path which is based on agent information. At the second level of the grade approach we try to achieve maximum bandwidth for routing using the following maximization technique,

Maximize $\sum_{j \in N(i)}^{n} l_{ij}$
Subject to $\sum_{j \in N(i)}^{n} P(l_j) = 1$
Over $0 \leq l_j \leq 1$ for all j

Here $l_{ij}$ represent the link capacity from node *i* to node *j*. We try to get the link which possesses maximum bandwidth. This is arrived by $P(l_j)$ which represent the probabilistic bandwidth value for a given node where its neighboring node are of high bandwidth.

## V. SIMULATION RESULTS

The initial configuration of simulation model consists of a randomly distributed set of nodes, in a geographical area. They are capable of getting connected through links which could be created on demand. The user enters the source and destination node for which a quadrant is drawn with source as the centre. With this structure, depending on the destination node selection of the path is carried out. The quadrant structure reduces the search space nearly to one-fourth depending on the source location. Here the agent evaluates all the nodes in the quadrant where destination node belongs and performs level-1 grading. The level-2 grading is performed using ABC algorithm. Table. 1 is the comparative result of graded ABC and non-graded ABC. It shows that the route length obtained by grade ABC is less than route length obtained by non-graded ABC. The column two of the table shows how the search space has reduced due to the application of the quadrant design in graded ABC.

TABLE I. RESULT ANALYSIS

| Total nodes | ABC with quality grading of nodes (A2) | | ABC without quality grading of nodes (A1) | |
|---|---|---|---|---|
| | Number of nodes selected | Route length | Number of nodes selected | Route length |
| 15 | 10 | 3 | 15 | 5 |
| 16 | 12 | 4 | 16 | 4 |
| 32 | 18 | 5 | 32 | 5 |
| 64 | 38 | 5 | 64 | 6 |
| 128 | 61 | 5 | 128 | 6 |

Here for implementing ABC, the number of onlooker bees is chosen based on the number of neighbors the source has. The employed bees now search for the best nectar among the available neighbors for each one of them. This search is continued until the destination is reached. Now among the different paths which exist between the source and destination one complete path is selected based on the bandwidth estimation of the links. Since here the link selected needs to be above a threshold value, there is less likely scenario for congestion to occur. Table. 2 shows the computation time obtained after applying ABC on graded and non-graded surface.

TABLE II. COMPUTATION TIME

| A2-Computation Time | A1- Computation Time |
|---|---|
| 3 | 5 |
| 4 | 6 |
| 5 | 7 |
| 5.5 | 7.5 |
| 6 | 8 |

Figure 4, depicts the graph for the throughput and traffic intensity determined in both graded and non graded network. It is seen that the traffic intensity in grade ABC is less than non-graded network. Also the throughput was more or less better in graded ABC than in non-graded ABC.

The paper here presented the optimal path determination using the bio inspired artificial bee colony approach for graded network. The approach showed a good performance in terms of convergence rate as the path search took place only in one fourth of the search area. A quality path with sufficient bandwidth was determined. Here we have shown that without a pre-defined message path we are able to obtain the paths. The routers are able to dynamically select the forward nodes and propagate the messages successfully to the destination. The result showed the traffic intensity was low in graded ABC as against to non-graded ABC.

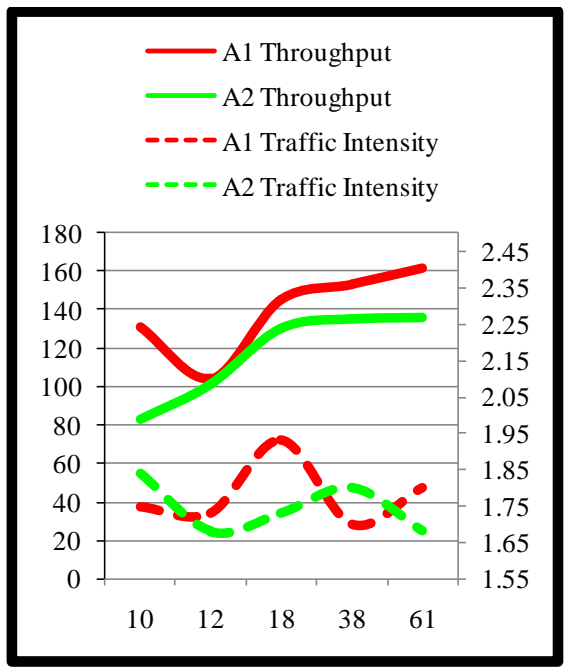

Figure 4. Graph plot for realizing through and traffic intensity for graded ABC and non-graded ABC

However the simulation has its own limitation. Currently only few hundred nodes were used to test and evaluate the principles of the model. Future work involves expanding the simulation to higher number of nodes realizing near real time network scenarios.

The simulation setup shall be modified to a zone based structure where there shall be a zone coordinator which will be responsible improving scalability and for achieving parallelism.